\documentclass[11pt,twoside]{article}
\usepackage{asp2010}

\newcommand\feh{[{\rm Fe/H}]}
\newcommand\vR{\rm{v_R}}
\newcommand\vphi{\rm{v_\phi}}
\newcommand\vz{\rm{v_z}}
\newcommand\dispvr{\sigma_{\rm{R}}}
\newcommand\sm{{\rm M}_\odot}
\newcommand\kms{{\rm \,km\,s^{-1}}}

\resetcounters
\begin{document}

\title{A small slice of the Milky Way disk in SDSS}
\author{Martin C. Smith$^{1,2}$
\affil{$^1$Kavli Institute for Astronomy and Astrophysics, Peking
  University, Beijing 100871, China; msmith@pku.edu.cn}
\affil{$^2$National Astronomical Observatories, Chinese Academy of Sciences, Beijing 100012, China}}

\begin{abstract}
The present-day state of the Milky Way disk can tell us much about the
history of our Galaxy and provide insights into its formation.
We have constructed a high-precision catalogue of disk
stars using data from the Sloan Digital Sky Survey (SDSS) and use these
stars to probe the heating history as well as investigating the
detailed phase-space distribution. We also show how this sample can be
used to probe the global properties of the Milky Way disk, employing
the Jeans equations to provide a simple model of the potential close
to the disk. Our model is in excellent agreement with others in the
literature and provides an indication that the disk, rather than the halo,
dominates the circular speed at the solar neighborhood.
The work presented in these proceedings has been published as
``Slicing and dicing the Milky Way disc in SDSS'' \citep{Sm12}.
\end{abstract}

\section{Introduction and sample construction}

Large surveys such as SDSS are beginning to enable detailed studies to
be made of the properties of the Milky Way. It is now possible to
construct high-precision catalogues of stars with full 3D kinematics
and from these analyse the structure and evolution of our
Galaxy. In the following proceedings I summarise our study of the
disk of the Milky Way \citep[presented in full in][]{Sm12},
highlighting two of the most important results.

We use data from the 7th public data release of the SDSS,
concentrating on the 250 square degree Stripe 82 region for
which it is possible to obtain high precision photometry and proper
motions \citep{Br08}. This catalogue has been exploited in a variety
of works, covering white dwarfs \citep{Vi07}, halo kinematics
\citep{Sm09,Sm09b} and RR Lyrae in the outer halo \citep{Wa09}.

We take all dwarf stars with good quality spectra \citep{Le08},
estimating distances using a slightly modified version of the
\citet{Iv08} photometric parallax relation. Our modification concerns
the turn-off correction \citep[equation A6 of][]{Iv08}, which we argue
is not applicable to disk stars \citep[see the Appendix of][]{Sm12}. 
We then calculate the three-dimensional velocities for our sample,
propagating all uncertainties, and obtain final errors of around 20 to
30 $\kms$ for each component of the velocity. 

As we are interested in the kinematics of the disk as a function of
height from the plane, we restrict ourselves to $7 \le \rm{R} \le
9$. We also split the data into three ranges in metallicity ($-1.5
\le \feh \le -0.8, -0.8 \le \feh \le -0.5$ and $-0.5 \le \feh$), and
then for each metallicity bin we further divide the data into four
ranges in z out to a maximum distance of 2 kpc. The stars are equally
divided between the four distance bins, resulting in in around $500$
to $800$ stars per bin from our full sample of 7280 stars. This
binning is necessary in order to avoid having to model the SDSS
spectroscopic selection function.

Once we have divided up our stars, we then proceed to fit the
distribution of kinematics using maximum likelihood methods
(incorporating the uncertainties on the individual velocities). 
Both $\vR$ and $\vz$ are fit using Gaussian distributions, but since
it is well known that the distribution of $\vphi$ for the disk is
highly skewed and non-Gaussian we adopt an asymmetric functional form
\citep[equation (15) of][]{Cu94}. When carrying out this analysis it
is crucial to remove contamination from halo stars, particularly for
the most metal-poor bins. We do this by including a non-rotating
Gaussian component into the velocity fits, where the level of
contamination (and hence amplitude of the Gaussian) is assumed to be
twice the number of counter-rotating stars. The halo dispersions are
fixed according to the values found in \citet{Sm09}.

\section{The rotation lag of the disk and the asymmetric drift}

\begin{figure}
\centering
\includegraphics[width=4.6cm]{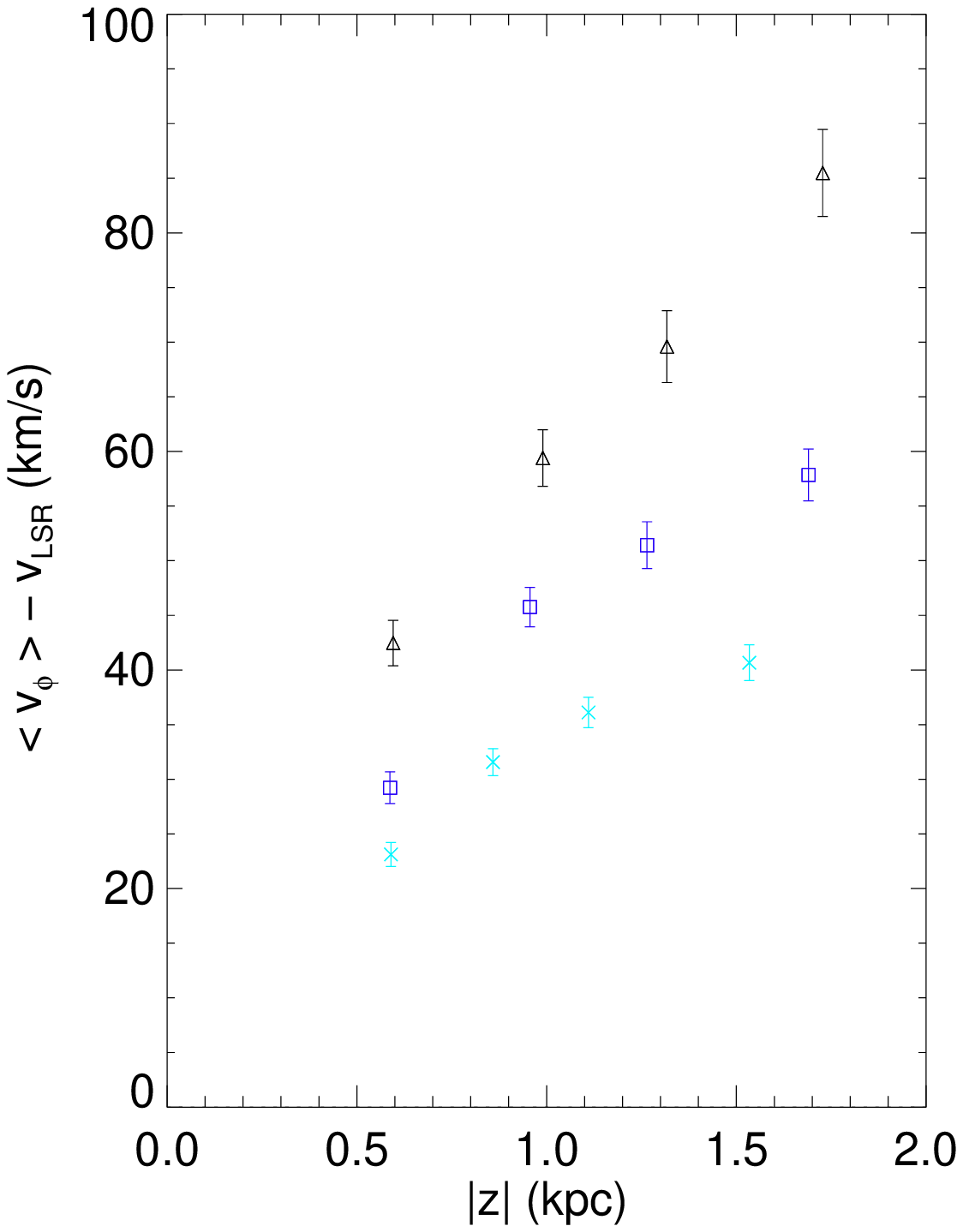}
\includegraphics[width=4.6cm]{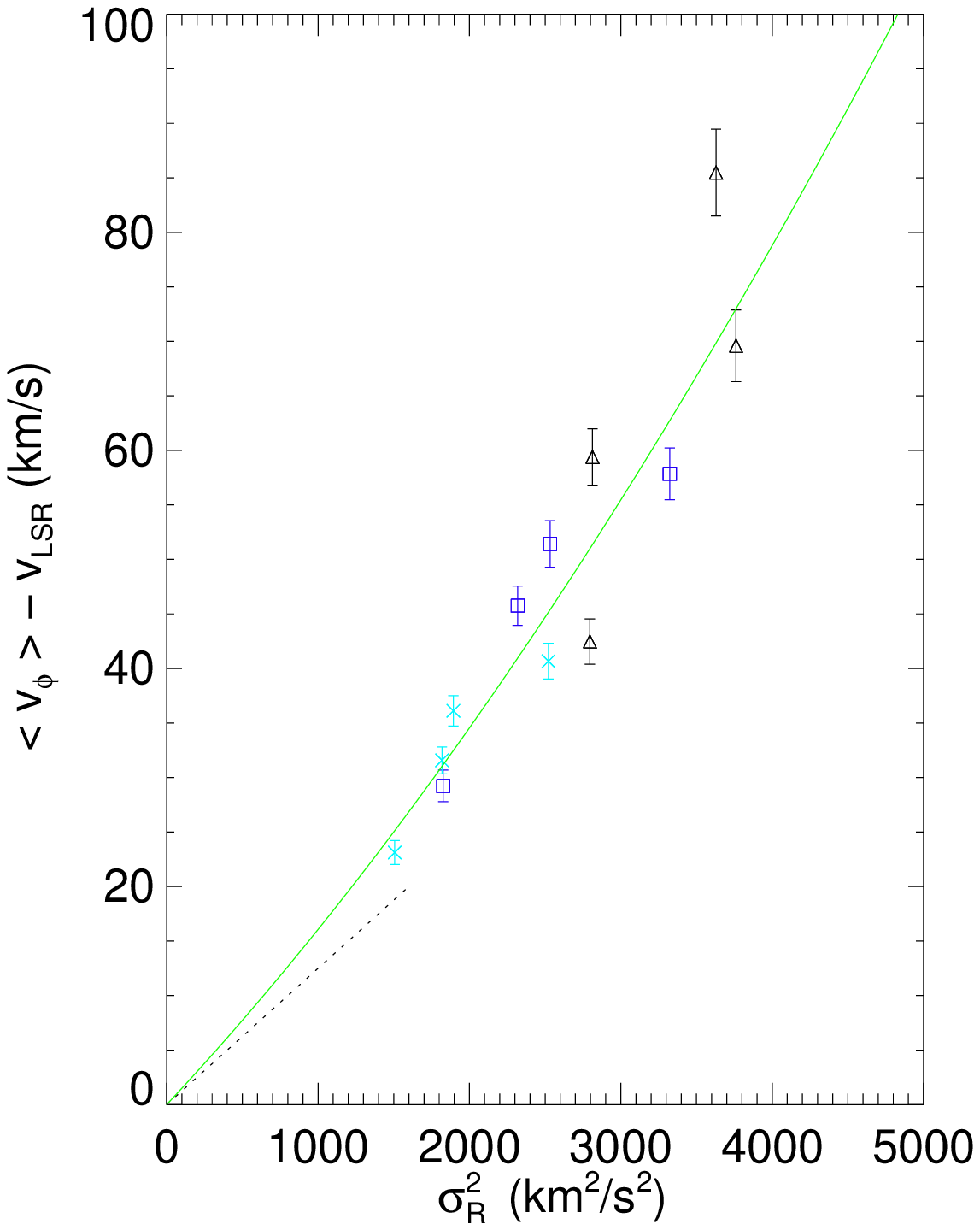}
\caption{The rotational lag, plotted against z (left) and against
  the radial velocity dispersion (right). The triangles, squares and
  crosses correspond to metal-poor, intermediate-metallicity and
  metal-rich populations, respectively. The dotted line in the right
  panel corresponds to the solar-neighborhood relation from \citet{DB}
  and the solid line denotes an empirical fit with the lag equal to
  $0.0149 \dispvr^2 + 1.21 \times 10 ^{-6} \dispvr^4$.
  Figure taken from \citet{Sm12}.}
\label{fig:lag}
\end{figure}

A comprehensive analysis of the results are presented in
\citet{Sm12}. Here we concentrate on the most important properties,
starting with rotation lag. We are able to trace the lag
(Fig. \ref{fig:lag}) to a couple of kpc outside the plane and find
that it follows the well-known asymmetric drift relation. The hotter
(metal-poor) stars exhibit greater lag than the colder (metal-rich)
populations, resulting in a clear correlation between lag and
metallicity. The gradient of the lag with respect to $|$z$|$ varies
from around 15 to 40 $\kms{\rm kpc}^{-1}$, depending on metallicity.
Fig. \ref{fig:lag} also shows how the lag is correlated with
$\dispvr^2$. For the solar-neighbourhood it is know that this is a
linear relation \citep{DB}, but as we can see from our figure this is
no longer true once we move beyond 0.5 kpc. However, it can be seen
that there is still a relatively tight correlation with $\dispvr^2$,
which is independent of metallicity.

\section{Constraining the gravitational potential of the disk}

\begin{figure}
\begin{center}
\includegraphics[width=5.55cm]{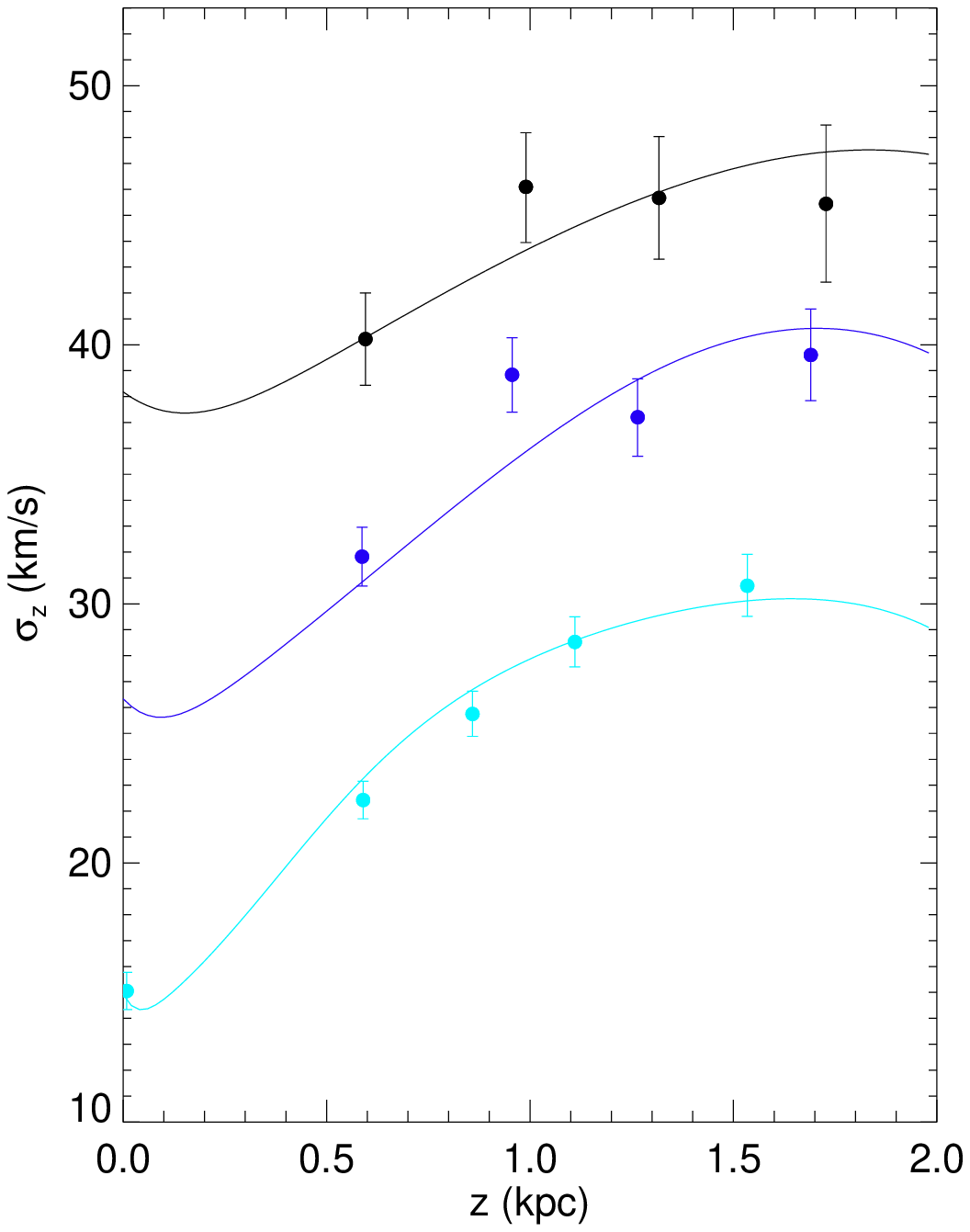}
\includegraphics[width=5.35cm]{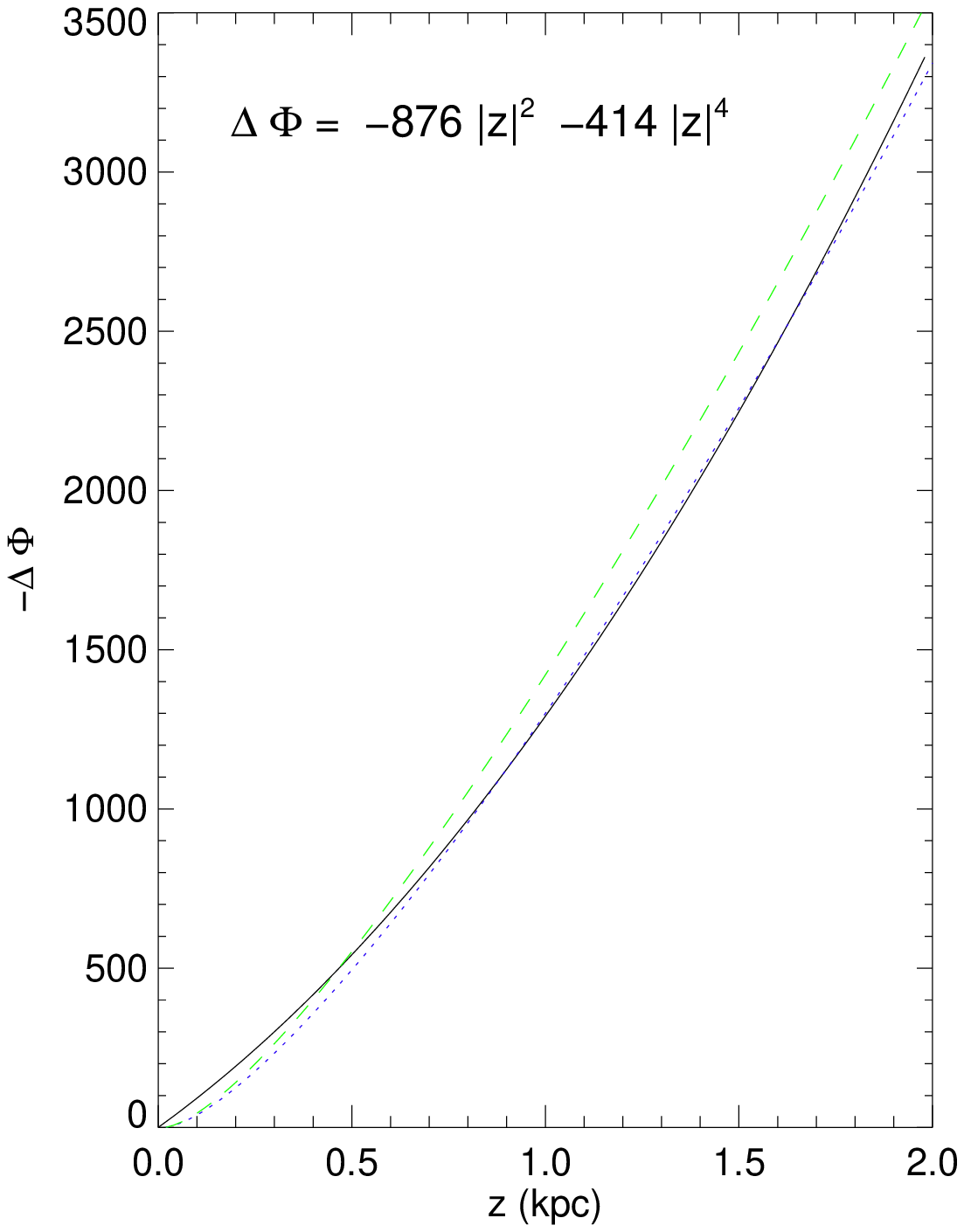}
\end{center}
\caption{The results of our modeling of the potential of
  the Galactic disk. The left panel shows the vertical velocity
  dispersion profiles (metal-poor - top; metal-rich - bottom), along
  with the corresponding profiles for the
  potential (solid lines). The right panel shows the potential
  resulting from this model. For the purposes of comparison we have
  included models for the potential taken from \citeauthor{DB98b}
  (1998 -- Model 1, dotted; Model 4, dashed). Figure adapted from
  \citet{Sm12}.}
\label{fig:potential}
\end{figure}

As has been shown by many authors, most notably in the seminal work of
\citet{Ku91}, the vertical velocity dispersion profile can be used to
constrain the gravitational potential of the disk. We show the results
derived from our data in Fig. \ref{fig:potential}, where we have
included an additional solar-neighbourhood metal-rich data point
derived from the Geneva-Copenhagen survey \citep{No04}. We take a
simplified two-parameter model for the potential, consisting of an
infinite razor-thin sheet embedded in a constant background. Our
best-fit model estimates that the mass density of these two components
are $32.5\: \sm {\rm pc}^{-2}$ and $0.015\:\sm {\rm pc}^{-3}$,
respectively.

As can be seen from Fig. \ref{fig:potential}, our measurement of the
potential is in good agreement with existing models, especially Model
1 of \citet{DB98b}. This is the least halo-dominated model of
\citet{DB98b}, with the disk (which has a relatively short
scale-length) dominating the circular speed at the solar
neighborhood. Interested readers should consult section 2.7 of
\citet{BT} for a detailed comparison of these models.

Note that this background mass density is close to the 
$0.014\:\sm {\rm pc}^{-3}$ predicted using isothermal spherical halo
models \citep[equation 4.279 of][]{BT}. If we assume our background
mass represents the dark halo, it corresponds to a local dark matter
density of 0.57 ${\rm GeV\;cm^{-3}}$, which is noticeably larger than
the canonical value of 0.30 ${\rm GeV\;cm^{-3}}$ typically assumed
\citep[e.g.][]{Ju96}.
Our analysis adds still more
weight to the argument the local halo density may be substantially
underestimated by this generally accepted value.
Perhaps more robust than the local mass density is the surface mass
density. By integrating our mass distribution we obtain a total
surface mass density of $\Sigma_{1.1\rm kpc} = 66 \: \sm {\rm pc}^{-2}$, which 
agrees well with the classical value of $71 \pm 6 \:\sm {\rm pc}^{-2}$
from \citet{Ku91}. If we integrate beyond 1.1 kpc, we find
$\Sigma_{2\rm kpc} = 94 \: \sm {\rm pc}^{-2}$ and 
$\Sigma_{4\rm kpc} = 155 \: \sm {\rm pc}^{-2}$.

\section{Conclusion}

Our results \cite[presented in full in][]{Sm12}
can be used to address the global properties of the disk
and also the heating processes that have shaped its evolution. In
order to fully understand the nature of disk heating and to
disentangle the contributions from various mechanisms, one needs to go
beyond the work presented here.
The most crucial improvement will be the ability to make direct
estimates for stellar ages, rather than relying on correlations with
metallicity. One aspect that will help us in this effort is by folding
in measurements of alpha-element abundances that are now being
determined routinely for vast numbers of stars \citep[e.g.][]{Le11}.

{\bf Question:} (Nicolas Martin) Can you develop your model to
test for the presence of a dark disk \citep[e.g.][]{Re08}?
{\bf Reply:} Due to the limitations of the data that we have used
there are a number of uncertainties in our modelling. As a consequence
of these limitations, I think it would be difficult to include extra
parameters.

\acknowledgements MCS acknowledges financial support from the Peking
University One Hundred Talent Fund (985) and NSFC grants 11043005 and
11010022. This work was also supported by the European Science
Foundation (ESF) for the activity entitled 'Gaia Research for European
Astronomy Training'.

\end{document}